\documentclass[preprint,3p,times,onecolumn]{elsarticle}

\usepackage[cmex10]{amsmath}
\usepackage{amssymb}
\usepackage{amsmath}
\usepackage{mathtools}
\usepackage{accents}
\usepackage{epsfig}
\usepackage{epstopdf}
\usepackage{mdwmath}
\usepackage{mdwtab}
\usepackage{stfloats}
\fnbelowfloat
\usepackage{IEEEtrantools}
\usepackage{color}

\usepackage{array}

\usepackage{mdwmath}
\usepackage{mdwtab}

\biboptions{sort&compress}

\journal{Signal Processing}

\makeatletter
\def\widebar{\accentset{{\cc@style\underline{\mskip15mu}}}}
\def\Widebar{\accentset{{\cc@style\underline{\mskip8mu}}}}
\makeatother

\newcommand{\ACal}{\mathcal{A}}
\newcommand{\Ab}{\mathbf{A}}
\newcommand{\ab}{\mathbf{a}}

\newcommand{\bb}{\mathbf{b}}

\newcommand{\eb}{\mathbf{e}}

\newcommand{\Xb}{\mathbf{X}}
\newcommand{\Xbt}{\widetilde{\Xb}}

\newcommand{\Xbr}{\mathbf{X}_{(r)}}
\newcommand{\Xbrc}{\mathbf{X}_{(-r)}}

\newcommand{\Yb}{\mathbf{Y}}
\newcommand{\xb}{\mathbf{x}}

\newcommand{\xbz}{\xb_0}
\newcommand{\xbd}{\xb^{\downarrow}}

\newcommand{\zb}{\mathbf{z}}
\newcommand{\ub}{\mathbf{u}}
\newcommand{\vb}{\mathbf{v}}
\newcommand{\Zerb}{\mathbf{0}}

\newcommand{\Rbb}{\mathbb{R}}
\newcommand{\matdim}{\Rbb^{n_1 \times n_2}}
\newcommand{\LOp}{\ACal: \Rbb^{n_1 \times n_2} \rightarrow \Rbb^m}

\newcommand{\sigminp}{\sigma_{\min,p}}

\newcommand{\vecdim}{\mathbb{R}^{n \times m}}

\newcommand{\xbt}{\widetilde{\xb}}

\newcommand{\xbi}{\xb_{I}}
\newcommand{\xbic}{\xb_{\bar{I}}}

\DeclareMathOperator{\rank}{rank}
\DeclareMathOperator{\spark}{spark}

\DeclareMathOperator{\nullS}{null}

\definecolor{purple}{rgb}{0.625, 0.125, 0.905}
\definecolor{DarkGreen}{rgb}{0.0, 0.5, 0.0}

\def\ROne_Col{black}
\def\RTre_Col{black}
\def\MinEdt{black}

\begin{document}

\begin{frontmatter}



\title{Upper Bounds on the Error of Sparse Vector and Low-Rank Matrix Recovery\tnoteref{t1}}
\tnotetext[t1]{This work was supported by the ACCESS Linnaeus Centre, KTH (Royal Institute of Technology), the Strategic Research Area ICT-TNG program, and the Swedish Research Council under contract 621-2011-5847.}



\author[add1]{Mohammadreza~Malek-Mohammadi\corref{cor1}}
\ead{mohamma@kth.se}
\author[add1]{Cristian R.~Rojas}
\ead{crro@kth.se}
\author[add1]{Magnus~Jansson}
\ead{janssonm@kth.se}
\author[add2]{Massoud~Babaie-Zadeh}
\ead{mbzadeh@yahoo.com}

\cortext[cor1]{Corresponding author.}

\address[add1]{ACCESS Linnaeus Centre, KTH, Stockholm, 10044, Sweden}
\address[add2]{Electrical Engineering Department, Sharif University of Technology, Tehran 1458889694}

\begin{abstract}
Suppose that a solution $\xbt$ to an underdetermined linear system $\bb = \Ab \xb$ is given. $\xbt$ is approximately sparse meaning that it has a few large components compared to other small entries. However, the total number of nonzero components of $\xbt$ is large enough to violate any condition for the uniqueness of the sparsest solution. On the other hand, 
if only the dominant components are considered, then it will satisfy the uniqueness conditions. One intuitively expects that $\xbt$ should not be far from the true sparse solution $\xbz$. We show that this intuition is the case by providing an upper bound on $\| \xbt - \xbz\|$ which is a function of the magnitudes of small components of $\xbt$ but independent from $\xbz$. 
This result is extended to the case that $\bb$ is perturbed by noise. Additionally, we generalize the upper bounds to the low-rank matrix recovery problem.
\end{abstract}


\end{frontmatter}


\section{Introduction} \label{sec:Intro}
%
%
%
%
Let 
$\xb_0 \in \Rbb^{m}$ denote a sparse solution of an underdetermined system of linear equations
\begin{equation} \label{USLE}
\bb = \Ab \xb
\end{equation}
in which 
$\bb \in \Rbb^{n}$ and 
$\Ab \in \vecdim, m > n$. Suppose that $\| \xb_0 \|_0 = k$, where $\| \xb_0 \|_0$ designates the number of nonzero components or the $\ell_0$ norm of $\xb_0$. Further, let $\spark(\Ab)$ represent the spark of $\Ab$, defined as the minimum number of columns of $\Ab$ which are linearly dependent, and let $\delta_{2k}(\Ab)$ denote the restricted isometry constant of order $2k$ for the matrix $\Ab$ \cite{Cand08}. It is well known that if $k < \spark(\Ab) / 2$ or $\delta_{2k}(\Ab) < 1$, then $\xb_0$ is the unique sparsest solution of the above set of equations \cite{Cand08,DonoE03}. 

When the sparsest solution of \eqref{USLE} is sought, one needs to solve
\begin{equation} \label{l0min}
\min_{\xb} \|\xb\|_0  \quad \text{subject to} \quad \Ab\xb=\bb.
\end{equation}
However, the above program is generally NP-hard \cite{Nata95} and becomes very intractable when the dimensions of the problem increase. Since finding the sparse solution of \eqref{USLE} has many applications in various fields of science and engineering (cf. \cite{CandW08} for a comprehensive list of applications), many practical alternatives for \eqref{l0min} have been proposed \cite{CandRT06,Trop04,MohiBJ09,MaleKBJR15}. If the solution obtained by these algorithms satisfies one of the above sufficient conditions, then, assuredly, this solution is the sparsest one.

Now, consider the case that the solution given by an algorithm is only approximately sparse meaning that it has some dominant components, while other components are very small but not equal to zero. If the total number of nonzero components is large such that neither of the mentioned conditions hold, it is not clear whether this solution is close to the true sparse solution or not. However, intuitively, one expects that if the number of effective components is small, then the obtained solution should not be far away from the true solution. Immediately, the following questions may be raised. Is this solution still close to the unique sparse solution of 
$\bb = \Ab \xb$? Is it possible in this case to establish a bound on the error of finding $\xb_0$ without knowing $\xb_0$? Similar questions can be asked when there is error or noise in \eqref{USLE}. Taking the noise into account, \eqref{USLE} is updated to
\begin{equation} \label{USLENoisy}
\bb = \Ab \xb + \eb,
\end{equation}
where 
$\eb$ is the vector of noise or error. In this setting, to estimate $\xb_0$ given 
$\bb$ and $\Ab$, the equality constraint in \eqref{l0min} is relaxed, and the following optimization problem should be solved:
\begin{equation} \label{l0minNoisy}
\min_{\xb} \|\xb\|_0  \quad \text{subject to} \quad \| \Ab\xb - \bb\| \leq \epsilon,
\end{equation}
where 
$\epsilon \geq \| \eb \|$ is some constant and $\| \cdot \|$ designates the $\ell_2$ norm.

The answers to the above questions were firstly given in \cite{BabaJM11}. Let $\xbt$ denote the output of an algorithm to find or estimate $\xb_0$ from \eqref{USLE} or \eqref{USLENoisy}. Particularly, \cite{BabaJM11} provides two upper bounds on the error $\| \xb_0 - \xbt \|$. The first one is rather simple to compute but turns out to be loose. On the other hand, while the second bound is tight, generally, it is much more complicated to compute. 

Herein, in the spirit of the loose bound in \cite{BabaJM11}, we provide a better bound which is based on the same parameter of the matrix $\Ab$, but it is \emph{strictly tighter} than the loose bound in \cite{BabaJM11}. Moreover, our proposed bound is obtained in a much simpler way with a \emph{shorter} algebraic manipulation. The proposed bound is extended to the noisy setting defined in \eqref{USLENoisy}. 
Furthermore, these results are also generalized to the problem of low-rank matrix recovery from compressed linear measurements \cite{RechFP10}.

The bounds introduced in this paper can be used in analyzing the performance of algorithms in sparse vector and low-rank matrix recovery, especially those algorithms that provide approximately sparse or low-rank solutions such as \cite{MohiBJ09} and \cite{MaleBAJ14,MaleBS14}. {\color{\RTre_Col} Other algorithms, under some conditions, can also benefit from the analysis presented in this paper. It is known that the solution obtained by some numerical solvers of basis pursuit \cite{ChenDS98}, like $\ell_1$-magic \cite{CandR05}, is not usually exactly sparse. In fact, due to limited numerical accuracy, the obtained solution has some very small nonzero entries. Our results can be used to find upper bounds on the $\ell_2$ norm of this kind of errors. Finally, when greedy algorithms \cite{Trop04} are used with an overestimated number of nonzero components of the true solution, our bound can be exploited to characterize the conditions under which the given solution is close to the true one.}
However, the bounds are obtained without any assumption on the recovery algorithm, and it is possible to improve them by exploiting properties of a specific algorithm. A similar upper bound on the error of sparse recovery in the noisy case has been proposed in \cite{GribFV06}. This upper bound, however, is only applicable when the given solution has a sparsity level{\color{\RTre_Col}, the number of nonzero components,} not greater than that of the true solution, while our bounds are obtained under the opposite assumption on the sparsity level of the given solution.


The rest of this paper is organized as follows. After introducing the notations used throughout the paper, in Section \ref{sec:Bounds}, we first present the upper bounds on the error of sparse vector recovery and, next, generalize them to the low-rank matrix recovery problem. Section \ref{sec:Proofs} is devoted to the proofs of the results in Section \ref{sec:Bounds}, followed by conclusions in Section \ref{sec:Con}.

\emph{Notations}: For a vector $\xb$, $\| \xb \|, \| \xb \|_1$, and $\| \xb \|_0$ denote the $\ell_2$, $\ell_1$, and the so-called $\ell_0$ norms, respectively. Moreover, $\xbd$ denotes a vector obtained by sorting the elements of $\xb$ in terms of magnitude in descending order, and $x_i$ designates the $i$th component of $\xb$. $\xb_{I}$ represents the subvector obtained from $\xb$ by keeping components indexed by the set $I$. A vector is called $k$-sparse if it has exactly $k$ nonzero components. For a matrix $\Ab$, $\ab_i$ denotes the $i$th column. Additionally, $\spark(\Ab)$ and $\nullS(\Ab)$ designate the minimum number of columns of $\Ab$ that are linearly dependent and the null space of $\Ab$, respectively. Similar to the vectors, $\Ab_{I}$ represents the submatrix of $\Ab$ obtained by keeping those columns indexed by $I$. It is always assumed that the singular values of matrices are sorted in descending order, and $\sigma_i(\Xb)$ denotes the $i$th largest singular value of $\Xb$. Let $\Xb = \sum_{i=1}^{q} \sigma_i \ub_i \vb_i^T$, where $q = \rank(\Xb)$, denote the singular value decomposition (SVD) of $\Xb$. $\Xbr = \sum_{i=1}^{r} \sigma_i \ub_i \vb_i^T$ represents a matrix obtained by keeping the $r$ first terms in the SVD of $\Xb$, and $\Xbrc = \Xb - \Xbr$. $\|\Xb\|_F$ denotes the Frobenius norm, and $\|\Xb\|_* \triangleq \sum_{i=1}^{q} \sigma_i(\Xb)$, in which $q = \rank(\Xb)$, stands for the nuclear norm.

\section{Upper Bounds} \label{sec:Bounds}
In this section, the upper bounds on the error of sparse vector and low-rank matrix recovery are presented.
\subsection{Sparse Vector Recovery} \label{SpRec}
Following the common practice in the literature of compressive sensing (CS), we refer to 
$\bb, \Ab$, and 
$\eb$ in \eqref{USLENoisy} as the measurement vector, sensing matrix, and noise vector, respectively. Before stating the results, we recall two definitions.
\newtheorem{Def1}{Definition}
\begin{Def1}[\hspace{-0.05em}\cite{Cand08}]
 For a matrix $\Ab \in \vecdim$ and all integers 
 $k \leq m$, the restricted isometry constant (RIC) of order $k$ is the smallest constant $\delta_k(\Ab)$ such that
 \begin{equation} \label{RIPDefInEq1}
 (1-\delta_{k}(\Ab)) \|\xb\|^2 \leq \|\Ab \xb\|^2 \leq (1+\delta_{k}(\Ab)) \|\xb\|^2
 \end{equation}
 holds for all vectors $\xb$ with sparsity at most $k$.
 \end{Def1}

 \newtheorem{Def2}[Def1]{Definition}
\begin{Def2}[\hspace{-0.05em}\cite{BabaJM11}]
For a matrix $\Ab \in \vecdim$, let $\sigminp(\Ab) > 0$ for $p \leq \spark(\Ab) - 1$ be the smallest singular value of all 
$\binom{m}{p}$ possible 
$n \times p$ submatrices of $\Ab$.
 \end{Def2}

The following theorem presents the upper bounds for both noisy and noiseless cases. We deliberately separate the noisy and noiseless cases in order to be able to provide a tighter bound in the noiseless setting.

\newtheorem{Thm1}{Theorem}
\begin{Thm1} \label{VecBound}
Let $\Ab \in \vecdim$, 
$m > n$, denote a sensing matrix. We have the following upper bounds.
\begin{itemize}
  \item Noiseless case: Suppose that $\xb_0$ is a $k$-sparse solution of 
  $\Ab \xb = \bb$, where $k < \spark(\Ab) / 2$. For all $\xbt$ solutions of 
  $\Ab \xb = \bb$ satisfying $\widetilde{x}^{\downarrow}_{k+1} \leq \alpha$,
      \begin{equation} \label{InEqVecSigNoNoise}
      \| \xb_0 - \xbt \|^2 \leq \Big( 1 + (m - 2k) \frac{\max_i \|\ab_i\|^2}{\sigma_{\min,2k}^2(\Ab)} \Big) (m - 2k) \alpha^2.
      \end{equation}
  \item Noisy case: Let $\xb_0$ be any arbitrary vector with $\| \xb_0 \|_0 = k < \spark(\Ab) / 2$, and let 
      $\bb = \Ab \xb_0 + \eb$, where 
      $\eb$ is noise with 
      $\| \eb \| \leq \epsilon$. For all $\xbt$ vectors satisfying 
      $\| \bb - \Ab \xbt \| \leq \Delta$ and $\widetilde{x}^{\downarrow}_{k+1} \leq \alpha$, the error $\| \xb_0 - \xbt \|$ is bounded by
      \begin{align} \label{InEqVecSigNoisy}
      \| \xb_0 - \xbt \| \leq & \Big( 1 + \sqrt{m - 2k}\frac{ \max_i \|\ab_i\|}{\sigma_{\min,2k}(\Ab)} \Big) \sqrt{m - 2k} \, \, \alpha \nonumber\\
       & + \frac{\Delta + \epsilon}{\sigma_{\min,2k}(\Ab)}.
      \end{align}
\end{itemize}
\end{Thm1}

In brief, the above bounds say that if we have a solution $\xbt$ that consists of $k$ large components, then this vector is not far from the sparse solution provided that $\sigma_{\min,2k}(\Ab)$ is not very small. 
In particular, the bound in \eqref{InEqVecSigNoNoise} vanishes when $\xbt$ is $k$-sparse, reducing to the well-known uniqueness theorem in \cite{DonoE03}. Moreover, notice that these bounds work uniformly for all sparse vectors $\xbz$ of sparsity level $k$; that is, they are independent from the position and magnitude of nonzero component of $\xbz$.

{\color{\MinEdt}\emph{Remark 1.}} The loose bounds in \cite[Theorems 2 \& 4]{BabaJM11} translated to our notations in the noiseless and noisy settings are
\begin{IEEEeqnarray}{rCl} \label{BZBound1}
\| \xbz - \xbt \| & \leq & \Big( 1 + \frac{1}{\sigma_{\min,2k}(\Ab)} \Big) m \alpha,\\
\| \xbz - \xbt \| & \leq & \Big( 1 + \frac{1}{\sigma_{\min,2k}(\Ab)} \Big) m \alpha + \frac{\Delta + \epsilon}{\sigma_{\min,2k}(\Ab)}. \label{BZBound2}
\end{IEEEeqnarray}
The bounds in \eqref{BZBound1} and \eqref{BZBound2} are applicable only if the sensing matrix has unit $\ell_2$ norm columns, whereas Theorem \ref{VecBound} is valid without this restriction. To compare our bounds in Theorem \ref{VecBound} to \eqref{BZBound1} and \eqref{BZBound2}, let $U$ denote the square root of the upper bound in \eqref{InEqVecSigNoNoise}. Substituting $\max_i \|\ab_i\|$ with 1 in $U$, one can write that
\begin{IEEEeqnarray*}{rCl}
U & = & \sqrt{\Big( 1 + \frac{m - 2k}{\sigma_{\min,2k}^2(\Ab)} \Big)(m - 2k)} \, \, \alpha\\
& < & \Big(1 +  \frac{\sqrt{m - 2k}}{\sigma_{\min,2k}(\Ab)} \Big)\sqrt{m - 2k} \, \, \alpha = U_2\\
& = & \Big(\frac{1}{\sqrt{m - 2k}} + \frac{1}{\sigma_{\min,2k}(\Ab)} \Big)(m - 2k) \alpha\\
& < & \Big( 1 + \frac{1}{\sigma_{\min,2k}(\Ab)}\Big) (m - 2k) \alpha\\
& < & \Big( 1 + \frac{1}{\sigma_{\min,2k}(\Ab)}\Big) m \alpha,
\end{IEEEeqnarray*}
where $U_2$ is the first term in the upper bound in \eqref{InEqVecSigNoisy} with $\max_i \|\ab_i\| = 1$. The above inequalities prove that the bounds \eqref{InEqVecSigNoNoise} and \eqref{InEqVecSigNoisy} are strictly tighter than the corresponding bounds in \cite{BabaJM11} formulated in \eqref{BZBound1} and \eqref{BZBound2}.

{\color{\MinEdt}\emph{Remark 2.}} {\color{\RTre_Col}In general, finding $\sigma_{\min,2k}(\Ab)$ is a combinatorial problem\footnote{{\color{\RTre_Col}Since one should calculate the singular values of all $\binom{m}{2k}$ possible $n \times 2k$ submatrices of $\Ab$.}}
 and NP-hard \cite{BabaJM11}. However, for a random matrix $\Ab$, under some conditions, the smallest singular value of all $n \times 2k$ submatrices is highly concentrated around a certain value. In particular, let $\Ab_{(2k)}$ denote any $n \times 2k$ submatrix of $\Ab$. If all the entries of $\Ab$ are independent and identically distributed (iid) from a normal distribution $\ensuremath{N(0,\frac{1}{n})}$ and $2k < n$, then for any $t > 0$, we have \cite{BabaJM11}
\begin{equation*}
p\Big\{ \sigma_{min}(\Ab_{(2k)}) < 1 - \sqrt{ \frac{2k}{n} } - t \Big\} \leq e^{-\frac{nt^2}{2}},
\end{equation*}
where $p\{\cdot\}$ and $\sigma_{min}(\cdot)$ denote the probability of the event described in the braces and the smallest singular value, respectively. This shows that when the dimensions of $\Ab$ increase, the smallest singular value of all $n \times 2k$ submatrices is equal to or larger than $1 - \sqrt{ \frac{2k}{n} }$ with very high probability. In line with this, for any matrix with iid entries from a zero-mean, $\frac{1}{n}$-variance distribution with a finite fourth-order moment, when $n,m \to \infty$ while $\frac{2k}{n} \to c$, $\sigma_{min}(\Ab_{(2k)})$ converges to $1 - \sqrt{c}$ almost surely \cite{BaiY93}.}

{\color{\MinEdt}\emph{Remark 3.}} {\color{\MinEdt} In addition to the above probabilistic values for $\sigma_{\min,2k}(\Ab)$,}
the bounds in Theorem \ref{VecBound} can be also stated in terms of $\delta_{2k}(\Ab)$ instead of $\sigma_{\min,2k}(\Ab)$. In fact,
\begin{equation*}
\sigma_{\min,2k}(\Ab) = \min_{\| \xb \|_0 \leq 2k} \frac{\| \Ab \xb \|}{\| \xb \|},
\end{equation*}
or $\| \Ab \xb \|^2 \geq \sigma_{\min,2k}^2(\Ab) \| \xb \|^2$ for all $\xb$ with sparsity at most $2k$. Since $\delta_{2k}(\Ab)$ in \eqref{RIPDefInEq1} is in such a way that both inequalities are satisfied, it can be concluded that $\sigma_{\min,2k}^2(\Ab) \geq 1 - \delta_{2k}(\Ab)$. Consequently, the following bounds, {\color{\ROne_Col}under the condition $\delta_{2k}(\Ab) < 1$}, are a reformulation of the bounds in Theorem \ref{VecBound} in terms of $\delta_{2k}(\Ab)$ which is frequently used in CS literature.
\begin{itemize}
\item Noiseless case:
\begin{equation*}
\| \xb_0 - \xbt \|^2 \leq \Big( 1 + (m - 2k) \frac{\max_i \|\ab_i\|^2}{1 - \delta_{2k}(\Ab)} \Big) (m - 2k) \alpha^2.
\end{equation*}
\item Noisy case:
\begin{align*}
\| \xb_0 - \xbt \| \leq & \Big( 1 + \sqrt{m - 2k} \frac{ \max_i \|\ab_i\|}{\sqrt{1 - \delta_{2k}(\Ab)}} \Big) \sqrt{m - 2k} \,\, \alpha \nonumber\\
       & + \frac{\Delta + \epsilon}{\sqrt{1 - \delta_{2k}(\Ab)}}.
\end{align*}
\end{itemize}

\subsection{Low-rank Matrix Recovery}
Recovery of a low-rank matrix from compressed linear measurements \cite{RechFP10} is the task of finding the low-rank matrix $\Xb_0 \in \matdim$ from underdetermined measurements $\bb = \ACal(\Xb_0)$ where $\bb \in \Rbb^{m}, \ACal:\matdim \to \Rbb^{m}$ is a linear operator, and $m < n_1 n_2$. In the presence of noise, the measurement model is changed to $\bb = \ACal(\Xb_0) + \eb$ where $\eb$ is the vector of noise.\footnote{The parameters $\bb,m,\eb,$ and $n$ (to be defined later on in this subsection) should not be confused with the similar parameters defined in Subsection \ref{SpRec}.} This recovery is a generalization of sparse vector recovery introduced in Section \ref{sec:Intro} to matrix variables. Consequently, the naive approach for recovering $\Xb_0$ from either noiseless or noisy measurements is
\begin{equation} \label{rankmin}
\min_{\Xb} \rank(\Xb)  \quad \text{subject to} \quad \| \ACal(\Xb) - \bb \| \leq \epsilon,
\end{equation}
where $\epsilon$ is some constant not less than $\| \eb \|$ in the noisy case and equal to 0 in the noiseless case.

In this subsection, we present upper bounds on the error of recovering or estimating low-rank matrices from noiseless and noisy measurements when the obtained solution is approximately low-rank. Similar to the vector case, a matrix is approximately low rank, if it is composed of a few dominant singular values, while its other singular values are very small. Before stating the results, first the definition of the RIC is recalled.

\newtheorem{Def3}[Def1]{Definition}
\begin{Def3}[\hspace{-0.05em}\cite{CandP11}]
For a linear operator $\LOp$ and all integers $r \leq \min(n_1,n_2)$, the RIC of order $r$ is the smallest constant $\delta_r(\ACal)$ such that
\begin{equation*}
(1-\delta_{r}(\ACal)) \|\Xb\|_F^2 \leq \|\ACal(\Xb)\|^2 \leq (1+\delta_{r}(\ACal)) \|\Xb  \|_F^2
\end{equation*}
holds for all matrices $\Xb$ with rank at most $r$.
\end{Def3}


\newtheorem{Thm2}[Thm1]{Theorem}
\begin{Thm2} \label{MatBound}
Let $\LOp, m < n_1 n_2,$ denote a linear operator, and let $n = \min(n_1,n_2)$. We have the following upper bounds.
\begin{itemize}
  \item Noiseless case: Suppose that $\Xb_0$ is a rank $r$ solution of $\bb = \ACal(\Xb)$. If $0 < \delta_{2r}(\ACal) < 1$, then, for all $\Xbt$ solutions of $\bb = \ACal(\Xb)$ satisfying $\sigma_{r+1}(\Xbt) \leq \alpha$,
      \begin{equation} \label{InEqMatSigNoNoise}
      \| \Xb_0 - \Xbt \|_F^2 \leq \Big( 1 + (n - 2r) \frac{1 + \delta_1(\ACal)}{1 - \delta_{2r}(\ACal)} \Big) (n - 2r) \alpha^2.
      \end{equation}
  \item Noisy case: Let $\Xb_0$ be any arbitrary matrix of rank $r$, and let $\bb = \ACal(\Xb_0) + \eb$, where $\eb$ is noise with $\| \eb \| \leq \epsilon$. If $0 < \delta_{2r}(\ACal) < 1$, then for all $\Xbt$ estimates of $\Xb_0$ satisfying $\| \bb - \ACal(\Xbt) \| \leq \Delta$ and $\sigma_{r+1}(\Xbt) \leq \alpha$, the error $\| \Xb_0 - \Xbt \|$ is bounded by
      \begin{align} \label{InEqMatSigNoisy}
      \| \Xb_0 - \Xbt \|_F \leq & \Bigg( 1 + \sqrt{(n - 2r)\frac{1 + \delta_1(\ACal)}{1 - \delta_{2r}(\ACal)}} \Bigg) \sqrt{n - 2r} \,\, \alpha \nonumber\\
       & + \frac{\Delta + \epsilon}{1 - \delta_{2r}(\ACal)}.
      \end{align}
\end{itemize}

\end{Thm2}

\section{Proofs of Results} \label{sec:Proofs}
\subsection{Proof of Theorem \ref{VecBound}}
We need the following lemmas.

\newtheorem{Lem1}{Lemma}
\begin{Lem1} \label{VecNoNoise}
Let $\Ab \in \vecdim$, 
$m > n$, be a sensing matrix. For every $\xb \in \nullS(\Ab)$ and any subset $I$ of 
$\{1,\cdots,m\}$ with cardinality 
$m - p$, where $ p \leq \spark(\Ab) -1$, we have that
\begin{equation} \label{InEqVecSig}
\| \xb \|^2 \leq \Big( 1 + (m - p) \frac{\max_i \|\ab_i\|^2}{\sigminp^2(\Ab)} \Big) \| \xbi \|^2.
\end{equation}
\begin{IEEEproof}
First, we notice that
\begin{IEEEeqnarray}{rCl}
\Big \| \sum_{i \in I} x_i \ab_i \Big \|^2 & \leq & \Big( \sum_{i \in I}  \| x_i \ab_i \| \Big)^2 = \Big( \sum_{i \in I} |x_i| \| \ab_i \| \Big)^2,\nonumber \\
& \leq & \max_i \| \ab_i \|^2 \Big( \sum_{i \in I}| x_{i} | \Big)^2, \nonumber \\
& = & \max_i \| \ab_i \|^2 \| \xbi \|_1^2, \nonumber \\
& \leq & (m - p) \max_i \| \ab_i \|^2 \| \xbi \|^2, \label{InEq1}
\end{IEEEeqnarray}
where, for the last inequality, we used $\forall \zb \in \Rbb^{l}, \| \zb \|_1^2 \leq l \| \zb \|^2$ \cite{HornJ90}. Next, from $\Ab \xb = \sum_{i \in I} x_i \ab_i + \sum_{i \notin I} x_i \ab_i = 0$, we get
\begin{equation} \label{InEq2}
\Big \| \sum_{i \in I} x_i \ab_i \Big \|^2 = \| \Ab_{\bar{I}} \xbic \|^2 \geq \sigminp^2(\Ab) \| \xbic \|^2,
\end{equation}
where 
$\bar{I} = \{1,\cdots,m\} \setminus I$. Combining inequalities \eqref{InEq1} and \eqref{InEq2} and using $\| \xb \|^2 = \| \xbi \|^2 + \| \xbic \|^2$ prove \eqref{InEqVecSig}.
{\color{\RTre_Col}Note that $p \leq \spark(\Ab) - 1$ implies that $\sigminp(\Ab) \neq 0$ and inequality \eqref{InEqVecSig} is not trivial.}
\end{IEEEproof}
\end{Lem1}

\newtheorem{Lem2}[Lem1]{Lemma}
\begin{Lem2} \label{VecNoisy}
Let $\Ab \in \vecdim$, 
$m > n$, be a sensing matrix. For every $\xb$ satisfying $\| \Ab \xb \| \leq \eta$ and every subset $I$ of 
$\{1,\cdots,m\}$ with cardinality 
$m - p$, where $ p \leq \spark(\Ab) -1$, we have that
\begin{equation} \label{InEqVecNoisySig}
\| \xb \| \leq \Big( 1 + \sqrt{m - p} \frac{\max_i \|\ab_i\|}{\sigminp(\Ab)} \Big) \| \xbi \| + \frac{\eta}{\sigminp(\Ab)}.
\end{equation}
\begin{IEEEproof}
Similar to the proof of Lemma \ref{VecNoNoise}, we have
\begin{equation}
\Big \| \sum_{i \in I} x_i \ab_i \Big \| \leq \sqrt{m - p} \max_i \| \ab_i \| \| \xbi \|.\label{InEq4}
\end{equation}
Furthermore, from $\Ab \xb = \sum_{i \in I} x_i \ab_i + \sum_{i \notin I} x_i \ab_i$, we get
\begin{IEEEeqnarray}{rCl} \label{InEq5}
\Big \| \sum_{i \in I} x_i \ab_i \Big \| & \geq & \| \Ab_{\bar{I}} \xbic \| - \| \Ab \xb \|,\nonumber \\
& \geq & \sigminp(\Ab) \| \xbic \| - \| \Ab \xb \|, \nonumber \\
& \geq & \sigminp(\Ab) \| \xbic \| - \eta.
\end{IEEEeqnarray}
Combining inequalities \eqref{InEq4} and \eqref{InEq5} leads to
\begin{equation*}
\sigminp(\Ab) \| \xbic \| \leq \sqrt{m - p} \max_i \|\ab_i\| \| \xbi \| + \eta
\end{equation*}
which is equivalent to
\begin{equation*}
\| \xbi \| + \| \xbic \| \leq \Big(1 + \sqrt{m - p}\frac{\max_i \|\ab_i\|}{\sigminp(\Ab)} \Big)  \| \xbi \| + \frac{\eta}{\sigminp(\Ab)}.
\end{equation*}
The above inequality together with
\begin{equation*}
\| \xb \| = \Bigg \| \begin{bmatrix} \xbi \\ \xbic \end{bmatrix} \Bigg \| \leq \Bigg \| \begin{bmatrix} \xbi \\ \Zerb \end{bmatrix} \Bigg \| + \Bigg \| \begin{bmatrix} \Zerb \\ \xbic \end{bmatrix} \Bigg \| = \| \xbi \| + \| \xbic \|,
\end{equation*}
where $\Zerb$ is a vector of zeros of appropriate length, proves \eqref{InEqVecNoisySig}.
\end{IEEEproof}
\end{Lem2}

\begin{IEEEproof}[Proof of Theorem \ref{VecBound}]
To prove \eqref{InEqVecSigNoNoise}, we first notice that because $\xb_0$ has $k$ nonzero components and $\widetilde{x}^{\downarrow}_{k+1} \leq \alpha$, $\xb = \xb_0 - \xbt$ has at most $2k$ components with magnitude larger than $\alpha$. Alternatively, $\xb$ possesses at least 
$m - 2k$ components with magnitude not greater than $\alpha$. Now, let $I$ denote a set of indexes of components of $\xb$ with magnitude less than or equal to $\alpha$ such that 
$|I| = m - 2k$. It is clear that 
$\| \xbi \|^2 \leq (m - 2k) \alpha^2$. Consequently, since $\xb \in \nullS(\Ab)$, we can apply Lemma \ref{VecNoNoise} to get
\begin{IEEEeqnarray*}{rCl}
\| \xb_0 - \xbt \|^2 & \leq & \Big( 1 + (m - 2k) \frac{\max_i \|\ab_i\|^2}{\sigma_{\min,2k}^2(\Ab)} \Big) \| \xbi \|^2, \\
& \leq & \Big( 1 + (m - 2k) \frac{\max_i \|\ab_i\|^2}{\sigma_{\min,2k}^2(\Ab)} \Big)(m - 2k) \alpha^2.
\end{IEEEeqnarray*}
For proving \eqref{InEqVecSigNoisy}, we start with
\begin{IEEEeqnarray}{rCl}
\| \Ab (\xb_0 - \xbt) \| & = & \| \bb - \Ab \xbt + \Ab \xb_0 - \bb \|,\nonumber\\
&  \leq & \| \bb - \Ab \xbt \| + \| \Ab \xb_0 - \bb \|,\nonumber\\
& \leq & \Delta + \epsilon. \label{InEqThm1_1}
\end{IEEEeqnarray}
Following the same reasoning as in the proof of \eqref{InEqVecSigNoNoise}, the application of Lemma \ref{VecNoisy} proves \eqref{InEqVecSigNoisy}.
\end{IEEEproof}

\subsection{Proof of Theorem \ref{MatBound}}

\newtheorem{Lem3}[Lem1]{Lemma}
\begin{Lem3} \label{MatNoNoise}
Let $\LOp, m < n_1 n_2,$ denote a linear operator. For every $r < n = \min(n_1,n_2)$ and every $\Xb \in \nullS(\ACal)$, if $0 < \delta_{r}(\ACal) < 1$, then
\begin{equation} \label{InEqLem3_0}
\| \Xb \|_F^2 \leq \Big( 1 + (n - r) \frac{1 + \delta_{1}(\ACal)}{1 - \delta_{r}(\ACal)} \Big) \| \Xbrc \|_F^2.
\end{equation}
\begin{IEEEproof}
Let $\Xb = \sum_{i=1}^{n} \sigma_i \ub_i \vb_i^T$ denote the SVD of $\Xb$. We can write that
\begin{IEEEeqnarray}{rCl}
\Big \| \ACal(\Xbrc) \Big \|^2 & = & \Big \| \ACal \Big ( \sum_{i=r+1}^n \sigma_i \ub_i \vb_i^T \Big) \Big\|^2,\nonumber\\
 & = & \Big \| \sum_{i=r+1}^n \sigma_i \ACal( \ub_i \vb_i^T ) \Big \|^2,\nonumber\\
& \leq &  \Big( \sum_{i=r+1}^n \sigma_i \big \| \ACal( \ub_i \vb_i^T ) \big \| \Big) ^2,\nonumber \\
& \overset{(a)}{\leq} &  \Big( \sum_{i=r+1}^n \sigma_i \sqrt{1 + \delta_{1}(\ACal)} \Big)^2,\nonumber \\
& = &  \big(1 + \delta_{1}(\ACal)\big) \big \| \Xbrc \big \|_*^2,\nonumber \\
& \overset{(b)}{\leq} &  (n - r) \big(1 + \delta_{1}(\ACal)\big) \big \| \Xbrc \big \|_F^2,\label{InEqLem3_1}
\end{IEEEeqnarray}
where (a) follows from the definition of the RIC and $\| \ub_i \vb_i^T \|_F = 1$ and for (b), we used the inequality $\| \Yb \|_* \leq \sqrt{\rank(\Yb)}\| \Yb \|_F$ \cite{HornJ90}.

Additionally, $\ACal(\Xb) = \ACal(\Xbr) + \ACal(\Xbrc) = \Zerb$ implies that
\begin{equation} \label{InEqLem3_2}
\big \| \ACal(\Xbrc) \big \|^2 = \big \| \ACal(\Xbr) \big \|^2 \geq \big(1 - \delta_{r}(\ACal) \big) \| \Xbr \|_F^2.
\end{equation}
Combining \eqref{InEqLem3_1} and \eqref{InEqLem3_2} together with $\| \Xb \|_F^2 = \| \Xbr \|_F^2 + \| \Xbrc \|_F^2$ leads to inequality \eqref{InEqLem3_0}.
\end{IEEEproof}
\end{Lem3}

\newtheorem{Lem4}[Lem1]{Lemma}
\begin{Lem4} \label{MatNoisy}
Let $\LOp, m < n_1 n_2$, denote a linear operator. For every $r < n = \min(n_1,n_2)$ and every $\Xb$ satisfying  $\| \ACal(\Xb) \| \leq \eta$, if $0 < \delta_{r}(\ACal) < 1$, then
\begin{IEEEeqnarray}{rCl}
\| \Xb \|_F & \leq & \Bigg( 1 + \sqrt{(n - r)\frac{1 + \delta_1(\ACal)}{1 - \delta_r(\ACal)}} \Bigg) \| \Xbrc \|_F \nonumber\\
& & + \frac{\eta}{\sqrt{1 - \delta_r(\ACal)}}. \label{InEqMatNoisy}
\end{IEEEeqnarray}
\begin{IEEEproof}
Inequality \eqref{InEqLem3_1} holds for every $\Xb$; thus, it is possible to write
\begin{equation}  \label{InEqLem4_1}
\| \ACal(\Xbrc) \| \leq \sqrt{(n - r)(1 + \delta_1(\ACal))} \| \Xbrc \|_F.
\end{equation}
Furthermore, applying the triangle inequality on $\ACal(\Xbrc) = \ACal(\Xb) - \ACal(\Xbr)$, one can obtain
\begin{IEEEeqnarray}{rCl} \label{InEqLem4_2}
\big \| \ACal(\Xbrc) \big \| & \geq & \big \| \ACal(\Xbr) \big \| - \big \| \ACal(\Xb) \big \|,\nonumber \\
& \geq & \sqrt{1 - \delta_r(\ACal)} \| \Xbr \|_F - \eta.
\end{IEEEeqnarray}
Combining inequalities \eqref{InEqLem4_1} and \eqref{InEqLem4_2} together with $\| \Xb \|_F \leq  \| \Xbr \|_F + \| \Xbrc \|_F$ gives inequality \eqref{InEqMatNoisy}.
\end{IEEEproof}
\end{Lem4}

\begin{IEEEproof}[Proof of Theorem \ref{MatBound}]
To prove \eqref{InEqMatSigNoNoise}, let us first define $\Xb = \Xb_0 - \Xbt$. According to \cite[Thmeorem 3.3.16]{HornJ91}, for any $1 \leq i,j \leq n$ and $i + j \leq n+1$,
\begin{equation*}
\sigma_{i+j-1}(\Xb) \leq \sigma_{i}(\Xb_0) + \sigma_{j}(\Xbt).
\end{equation*}
Substituting $i$ and $j$ with $r+1$ in the above inequality leads to
\begin{equation*}
\sigma_{2r+1}(\Xb) \leq \sigma_{r+1}(\Xb_0) + \sigma_{r+1}(\Xbt) \leq \alpha.
\end{equation*}
Consequently, Lemma \ref{MatNoNoise} implies that
\begin{IEEEeqnarray*}{rCl}
\| \Xb_0 - \Xbt \|_F^2 & \leq & \Big( 1 + (n - 2r) \frac{1 + \delta_1(\ACal)}{1 - \delta_{2r}(\ACal)} \Big) \| \Xb_{(-2r)} \|_F^2,\\
& \leq & \Big( 1 + (n - 2r) \frac{1 + \delta_1(\ACal)}{1 - \delta_{2r}(\ACal)} \Big) (n - 2r) \alpha^2.
\end{IEEEeqnarray*}
For proving \eqref{InEqMatSigNoisy}, we start with
\begin{IEEEeqnarray*}{rCl}
\| \Ab (\Xb_0 - \Xbt) \| & = & \| \bb - \ACal( \Xbt ) + \ACal( \Xb_0 ) - \bb \|,\nonumber\\
&  \leq & \Delta + \epsilon.
\end{IEEEeqnarray*}
Following the same reasoning as in the proof of \eqref{InEqMatSigNoNoise}, the application of Lemma \ref{MatNoisy} completes the proof.
\end{IEEEproof}


\section{Conclusion} \label{sec:Con}
In this paper, we proposed upper bounds on the error of sparse vector recovery from both noiseless or noisy measurements when the obtained solution is approximately sparse. While these bounds are based on the same parameters as in the loose bounds of \cite{BabaJM11}, they are strictly tighter. We further generalized them to the problem of low-rank matrix recovery, when the solution at hand to recover the true low-rank matrix is approximately low rank.

\section{Acknowledgement}
{\color{\MinEdt}The authors would like to thank the anonymous reviewers for their helpful comments.}

\bibliographystyle{elsarticle-num}
\bibliography{Refs}







\end{document}